\documentclass[epj]{svjour}
\usepackage{graphics}
\begin{document}
\title{Test of the Universal Rise of Hadronic Total Cross Sections at Super-high Energies}
\author{Muneyuki Ishida\inst{1} \and Keiji Igi\inst{2}% etc
}                     % Do not remove
\institute{Department of Physics, School of Science and Engineering, 
Meisei University, Hino, Tokyo 191-8506, Japan 
\and Theoretical Physics Laboratory, RIKEN, Wako, Saitama 351-0198, Japan}
\date{Received: date / Revised version: date}
\abstract{
Saturation of the Froissart-Martin unitarity bound that the total cross sections increase
like log$^2(s/s_0)$ appears to be confirmed. Due to this statement, the $B{\rm log}^2(s/s_0)$
was assumed to extend the universal rise of all the total hadronic cross sections to reduce the number
of adjustable parameters by the COMPETE Collaboration in the Particle Data Group (2006). 
Based on this assumption of parametrization, we test if the assumption 
on the universality of $B$ is justified
through investigations of the value of
$B$ for $\pi^\pm p(K^\pm p)$ and $\bar pp, pp$ scatterings. 
We search for the simultaneous best fit
to $\sigma_{\rm tot}$ and $\rho$ ratios, using a constraint 
from the FESR of the $P^\prime$ type for $\pi^\mp p$ scatterings and 
constraints which are free from unphysical regions for $\bar pp,pp$ and 
$K^\pm p$ scatterings.
By including rich informations of the low-energy scattering data owing to the use of FESR, 
the errors of $B$ parameters decreases especially 
for $\pi p$.
The resulting value of $B_{pp}$ is consistent with $B_{\pi p}$ 
within two standard deviation, which appears to support 
the universality hypothesis.
\PACS{
      {11.55.Hx}{ Sum rules }   \and
      {13.85.Lg}{Total cross sections }
     } % end of PACS codes
} %end of abstract
\maketitle
%
%\section{Introduction}
%\label{intro}
%Your text comes here. Separate text sections with
%\section{Section title}
%\label{sec:1}
%and \cite{RefJ}
%\subsection{Subsection title}
%\label{sec:2}
%as required. Don't forget to give each section
%and subsection a unique label (see Sect.~\ref{sec:1}).

\noindent \underline{\it Purpose of this Paper}

    It is well-known as the Froissart-Martin unitarity bound \cite{[1]} 
    that the increase of total cross sections is at most log$^2\nu$. 
    It had not been possible, however, to discriminate between asymptotic log $\nu$ 
    and log$^2\nu$ fits if one uses $\pi N$ high-energy data alone above 70 GeV.
    Therefore, we have proposed \cite{[2]} to use rich informations of $\pi p$ total 
    cross sections at low- and intermediate-energy regions through the finite-energy 
    sum rules ( FESR ) of the $P^\prime$ type \cite{[3]} as well as \cite{[4],[5]} 
    in addition to total cross sections, and have arrived at the conclusion that 
    log$^2\nu$ behavior is preferred, i. e., the Froissart-Martin bound \cite{[1]} is saturated. 
    Cudell et al., ( COMPETE Collab. ) \cite{[6]} have considered several classes of analytic parametrizations 
    of hadronic scattering amplitudes, and compared their predictions to all available forward data 
    ( $pp$, $\bar pp$, $\pi p$, $Kp$, $\gamma p$, $\gamma \gamma$, $\Sigma^- p$ ). 
    Although these parametrizations were very close for $\sqrt s \ge   9$ GeV, 
    it turned out that they differ markedly at low energy, 
    where log$^2 s$ enables one to extend the fit down to $\sqrt s = 4$ GeV\cite{[6]}.

    The statement that the log$^2 \nu$ behaviour is preferred have been confirmed 
    in \cite{[7]} and \cite{[8]}. 
    In Ref. \cite{[6]}, the $B$log$^2(s/s_0)$ was assumed to extend the universal rise of 
    all the total hadronic cross sections. 
    This resulted in reducing the number of adjustable parameters. 
    Recently, however, it was pointed out in Ref. \cite{[9]} that \cite{[7]}, \cite{[8]} gave different predictions 
    for the value of $B$ for $\pi N$ and $NN$, i.e., different predictions 
    at superhigh energies:  $\sigma_{\pi N}^{\rm as} > \sigma_{NN}^{\rm as}$\cite{[7]} 
    and $\sigma_{\pi N}^{\rm as} \sim 2/3\ \sigma_{NN}^{\rm as}$\cite{[8]}.

    The purpose of this article is to investigate the value of $B$ for 
    $\pi^\pm p(K^\pm p)$ and $\bar pp,pp$ cases 
    in order to check 
if the assumption on
the universality of the coefficient $B$ 
%in front of log$^2(s/s_0)$ terms
is justified.
    We search for the simultaneous best fit to  $\sigma_{\rm tot}$, 
    the total cross sections, and $\rho$, the ratios of the real to imaginary part of 
    the forward scattering amplitude, using a constraint 
from the FESR of the $P^\prime$ type for $\pi^\pm p$ scatterings and constraints
which are free from unphysical regions for $\bar pp,pp$ and 
$K^\pm p$ scatterings\cite{[10]}.\\

\noindent \underline{\it Total cross sections, $\rho$ ratios and constraints}

Let us consider the forward $\bar pp,pp$, $\pi^\mp p$ and $K^\mp p$ scatterings.
We take both the crossing-even and crossing-odd forward scattering amplitudes,
$F^{(+)}$ and $F^{(-)}$,  
defined by 
%\vspace*{-0.8cm}
\begin{eqnarray}
F^{(\pm )}(\nu ) &=& \frac{f^{\bar ap}(\nu )\pm f^{ap}(\nu )}{2}\ ;\label{eq1} \\ 
& & 
\begin{array}{l}
f^{\bar ap}(\nu )=F^{(+)}(\nu )+ F^{(-)}(\nu )\\
f^{ap}(\nu ) = F^{(+)}(\nu ) - F^{(-)}(\nu )
\end{array}\ ,
\label{eq1A}
\end{eqnarray}
where $(\bar a,a)=(\bar p,p)$, $(\pi^-,\pi^+)$, $(K^-,K^+)$, respectively.
We assume
%\vspace{-0.8cm}
\begin{eqnarray}
{\rm Im}F^{(+)}(\nu ) &=& {\rm Im}R(\nu )+ {\rm Im}F_{P^\prime}(\nu ) \nonumber\\
 = \frac{\nu}{m^2}  ( c_0  &+& \left. c_1 {\rm log}\frac{\nu }{m} 
     + c_2 {\rm log}^2\frac{\nu }{m}  \right)
    + \frac{\beta_{P^\prime}}{m} \left( \frac{\nu}{m}\right)^{\alpha_{P^\prime}}\ 
\label{eq2}\\
{\rm Im}F^{(-)}(\nu ) &=& {\rm Im}F_V(\nu ) 
    = \frac{\beta_V}{m} \left( \frac{\nu}{m}\right)^{\alpha_V}\ 
\label{eq3}
\end{eqnarray}
at high energies for $\nu > N$. 
%with high-energy parameters $c_2,c_1,c_0,\beta_{P^\prime}$ and $\beta_V$.
Here $m=M$(proton mass)$, m=\mu$(pion mass) and $m=m_K$(kaon mass) 
for $\bar p(p)p$, $\pi p$ and $Kp$ scatterings, respectively. 
The $\nu ,k$ are the incident $\bar p(p)$, $\pi$ and $K$ energies, momenta in the 
laboratory system, respectively.
Using the crossing-even$/$odd property, $F^{(\pm )}(-\nu )=\pm F^{(\pm )}(\nu )^*$, 
the real parts are given by\cite{[8],[2]}
%\vspace*{-0.8cm}
\begin{eqnarray}
{\rm Re}F^{(+)}(\nu ) &=& \frac{\pi \nu}{2m^2}\left( c_1 + 2 c_2 {\rm ln}\frac{\nu}{m} \right)\nonumber\\
 && -\frac{\beta_{P^\prime}}{m}\left(\frac{\nu}{m}\right)^{\alpha_{P^\prime}}
  {\rm cot}\frac{\pi\alpha_{P^\prime}}{2} + F^{(+)}(0)\ , \label{eq4}\\
{\rm Re}F^{(-)}(\nu ) &=& \frac{\beta_V}{m}\left(\frac{\nu}{m}\right)^{\alpha_V}
  {\rm tan}\frac{\pi\alpha_V}{2} \ . 
\label{eq5}
\end{eqnarray}
The total cross sections $\sigma_{\rm tot}^{\bar ap}$, $\sigma_{\rm tot}^{ap}$
and the $\rho$ ratios $\rho^{\bar ap}$, $\rho^{ap}$ are given by
%\vspace*{-0.8cm}
\begin{eqnarray}
{\rm Im} f^{\bar ap,ap}(\nu ) &=& 
\frac{k}{4\pi}\sigma_{\rm tot}^{\bar ap,ap};\ 
\rho^{\bar ap} =
\frac{{\rm Re}f^{\bar ap}}{{\rm Im}f^{\bar ap}},
\rho^{ap} =
\frac{{\rm Re}f^{ap}}{{\rm Im}f^{ap}}, \ \  \ 
\label{eq6}
\end{eqnarray}
respectively.

Defining $
\tilde F^{(+)}(\nu ) = F^{(+)}(\nu ) - R(\nu ) - F_{P^\prime}(\nu) - F^{(+)}(0) 
      \sim \nu^{\alpha (0)}\ (\alpha (0) < 0)$,
for large value of $\nu$,
we have obtained the FESR\cite{[2]} in the spirit of $P^\prime$ sum rule\cite{[3]}
%\vspace*{-0.8cm}
\begin{eqnarray}
{\rm Re} & \tilde F^{(+)}& (m)  \nonumber\\ 
 &=& \frac{2P}{\pi}\int_0^m \frac{\nu}{k^2}{\rm Im}F^{(+)}(\nu )d\nu 
+\frac{1}{2\pi^2} \int_0^{\overline{N}} \sigma^{(+)}_{\rm tot}(k)dk \nonumber\\
 & &  - \frac{2P}{\pi} \int_0^N \frac{\nu}{k^2} \left\{  
       {\rm Im}R(\nu )+{\rm Im}F_{P^\prime}(\nu ) \right\} d\nu\ ,
\label{eq7}
\end{eqnarray}
where $\overline{N} = {\sqrt{N^2-m^2}} \simeq N$. Let us call Eq.~(\ref{eq7}) as\\ 
FESR(1)(0--$\overline{N}$).\footnote{
This sum rule should hold if no singularities extend above $J=0$ except for the 
Pomeron and $P^\prime$.
So, we adopt this sum rule rather than the higher-moment sum rule.
} 

The Eq.~(\ref{eq7}) gives directly a constraint
for $\pi p$ scattering,
%\vspace*{-0.8cm} 
\begin{eqnarray}
\frac{2P}{\pi} \int_0^N  \frac{\nu}{k^2} && \left\{  
       {\rm Im}R(\nu )+{\rm Im}F_{P^\prime}(\nu ) \right\} d\nu \nonumber\\
   - {\rm Re}R(\mu )  &&  - {\rm Re}F_{P^\prime}(\mu ) \nonumber\\
   =   -{\rm Re}F^{(+)}(\mu )   &&  
+({\rm pole\ term}) + \frac{1}{2\pi^2} \int_0^{\overline{N}} 
\sigma^{(+)}_{\rm tot}(k)dk\ .\ \ \ 
\label{eq8}
\end{eqnarray}
For $\pi p$ scattering, RHS can be estimated with sufficient accuracy,
regarding Eq.~(\ref{eq8}) as an exact constraint\cite{[2]}:\\
Re$F^{(+)}(\mu )$ is represented by scattering lengths
and pole term comes only from nucleon.
%, of which strength is estimated precisely
%from the $\pi NN$ coupling constant $g_r$.  
The last term 
%integral of $\sigma_{\rm tot}^{(+)}$ 
is estimated from the rich data of 
experimental $\sigma_{\rm tot}^{\pi^\mp p}$.

On the other hand, Eq.~(\ref{eq7}) for $\bar p(p)p$ scattering suffers from the 
unphysical regions coming from boson poles below the $\bar pp$ threshold.
Reliable estimates, however, are difficult. Similarly in $Kp$ scattering, 
poles of $\Lambda ,\Sigma$ resonant states contribute below $K^- p$ threshold.
In ref.\cite{[10]}, we have presented a new constraint, 
called FESR(1)($\overline{N}_1$-$\overline{N}_2$), free from unphysical regions.
We consider Eq.~(\ref{eq8}) with $N=N_1$ and $N=N_2$ ($N_2>N_1$). 
Taking the difference between these two relations, we obtain the relation
%\vspace*{-0.8cm}
\begin{eqnarray}
\frac{2}{\pi}\int_{N_1}^{N_2} && \frac{\nu }{k^2} 
\left\{  {\rm Im}R(\nu )+ {\rm Im}F_{P^\prime}(\nu )
 \right\}  d\nu \nonumber\\
 && = \frac{1}{2\pi^2} \int_{\overline{N}_1}^{\overline{N}_2} 
\sigma^{(+)}_{\rm tot}(k )dk \ \ .\ \ \ \ \ \ \ 
\label{eq9}
\end{eqnarray}
The RHS can be estimated from the experimental 
$\sigma_{\rm tot}^{\bar pp,pp}$ and $\sigma_{\rm tot}^{K^\mp p}$ data,\footnote{
Practically it is estimated from the fit to $\sigma_{\rm tot}$ in 2.5GeV$\le k \le 100$GeV
through phenomenological formula. 
See, ref.\cite{[10]} for detail.  
}
regarding Eq.~(\ref{eq9}) as an exact constraint.\\
%This relation is called FESR(1)($\overline{N}_1$-$\overline{N}_2$). 
%It is free from the unphysical region. 

%used in the analysis 
%of $\bar pp,pp$ scattering.

\noindent \underline{\it The general approach}

The formula, Eqs.~(\ref{eq1})-(\ref{eq6}), and the constraints, 
Eqs.~(\ref{eq8}) and (\ref{eq9}), are our starting points.  
The $\sigma_{\rm tot}^{\bar ap,ap}$ and $\rho^{\bar ap,ap}$
are fitted simultaneously for respective processes of $\bar p(p)p$, $\pi p$, $Kp$ 
scatterings.
The high-energy parameters $c_2,c_1,c_0,\beta_{P^\prime}$, $\beta_V$ are treated as 
process-dependent, while $\alpha_{P^\prime}$ and $\alpha_V$ are fixed with common values
for every process.   
The FESR(1)
($\overline{N}_1$-$\overline{N}_2$)(Eq.~(\ref{eq9})) and 
FESR(1)(0-$\overline{N}$)(Eq.~(\ref{eq8})) give constraints between 
$c_2,c_1,c_0$ and $\beta_{P^\prime}$
for $\bar p(p)p,Kp$ and $\pi p$ scatterings, respectively. 
$F^{(+)}(0)$ is treated as an additional parameter, and 
the number of fitting parameters is 5 for each process.
The resulting $c_2$ are related to the $B$ parameters, 
defined by $\sigma \simeq B {\rm log}^2(s/s_0)+\cdots$, 
%Since $B$ is related to our parameter $c_2$ 
through the equation
\begin{eqnarray}  
B_{ap} &=& \frac{4\pi}{m^2}c_2\ {\rm where}\ m=M,\mu ,m_K\ {\rm for}\ a=p,\pi ,K,\ \ \ 
\label{eq10}
\end{eqnarray}
and we can test the universality of $B$ parameters for the relevant processes.\\

\begin{table*}
\caption{
Values of parameters and $\chi^2$ in the best fits 
in $(\alpha_{P^\prime},\alpha_V)=(0.500,0.497)$.
Both total $\chi^2$ and respective $\chi^2$ for each data with the number of data points 
are given.
The result of $\bar p(p)p$ scattering is in the 1st row,
$\pi p$ scattering in the 2nd row and $Kp$ scattering in the 3rd row. 
The errors are given only for $c_2$. 
The values of $\beta_{P^\prime}$ are obtained from FESR.
}
\begin{tabular}{c|cccccc|c|cccc}
 &  $c_2$ & $c_1$  & $c_0$ & $F^{(+)}(0)$ & $\beta_V$ & $\beta_{P^\prime}$
  & $\frac{\chi^2_{\rm tot}}{N_D-N_P}$
     & $\frac{\chi^{2,\sigma}_{\bar ap}}{N^\sigma_{\bar ap}}$
     & $\frac{\chi^{2,\rho}_{\bar ap}}{N^\rho_{\bar ap}}$
     & $\frac{\chi^{2,\sigma}_{ap}}{N^\sigma_{ap}}$
     & $\frac{\chi^{2,\rho}_{ap}}{N^\rho_{ap}}$ \\
\hline
$\bar pp,pp$ &  0.0520$\pm$0.0041 & -0.259 & 6.67 & 11.1 & 3.75 & 6.82 &
$\frac{157.3}{231-5}$ & $\frac{21.0}{51}$ & $\frac{15.4}{16}$ & $\frac{61.4}{94}$ & $\frac{59.5}{70}$\\ 
\hline
$\pi^\mp p$ & (140$\pm$14)$\cdot$10$^{-5}$
%0.00140$\pm$0.00014
 & -0.0153 & 0.132 & 0.291 & 0.0392 &  0.0875 & 
$\frac{42.3}{151-5}$ & $\frac{10.3}{84}$ & $\frac{11.7}{22}$ & $\frac{8.3}{37}$ & $\frac{12.0}{8}$\\
\hline
$K^\mp p$ & 0.0185$\pm$0.0103 & -0.142 & 1.20 & 1.86 & 0.573 & 0.234 &
$\frac{36.3}{111-5}$ & $\frac{17.9}{53}$ & $\frac{11.5}{13}$ & $\frac{5.9}{31}$ & $\frac{1.0}{14}$\\
\hline
\end{tabular}
\label{tab1}
\end{table*}

\noindent \underline{\it Result of the analyses}

The $\sigma_{\rm tot}^{\bar ap,ap}$ for $k\ge 20$ GeV and $\rho^{\bar ap,ap}$ for $k\ge 5$ GeV
are fitted sumultaneously.\footnote{In the actual analysis we fit data of Re$f$ 
instead of $\rho$. Re$f$ data are made from the original $\rho$ data multiplied by $\sigma_{\rm tot}$,
which is given by the fit in 
ref.\cite{[9]}.} 
We take two cases, $(\alpha_{P^\prime},\alpha_V)=(0.500,0.497)$ and
(0.542,0.496). These values are selected by considering the $\chi^2$ behaviours of the fit to 
$\bar pp,pp$ data: The total $\chi^2$ is almost independent of the input value of $\alpha_{P^\prime}$, while it is sensitive to the value of $\alpha_V$.
So we select two values of $\alpha_{P^\prime}$ 
as typical examples,\footnote{
Our $\alpha_{P^\prime}$ corresponds to $1-\eta_1$ in parametrization of COMPETE collab.\cite{[9]}.
$\alpha_{P^\prime}=0.542$ corresponds to their best fit value, $\eta_1=0.458$ . 
} 
while $\alpha_V$ is selected
from the minima of $\chi^2$. The $\chi^2$ takes its minimum
at $\alpha_V\sim 0.50$ independently of $\alpha_{P^\prime}$-value.

The FESR(1)(10-20GeV)(Eq.~(\ref{eq9})) for $\bar p(p)p,Kp$ and 
FESR(1)(0-20GeV)(Eq.~(\ref{eq8})) for $\pi p$ are given respectively by
\begin{eqnarray} 
 1.837(2.061) \beta_{P^\prime} && + 7.247 c_0 + 19.96 c_1 + 55.27 c_2 = 58.54     
\ \ \ \ \ \ \label{eq11}\\
 4.810(5.542) \beta_{P^\prime} && + 26.14 c_0 + 88.74 c_1 + 302.3 c_2 = 25.41 
\ \ \ \ \ \ \label{eq13}\\
 109.2(124.1) \beta_{P^\prime} && + 653.6 c_0 + 2591 c_1 + 10928 c_2 = 71.12
\ \ \ \ \ \ \label{eq12}
\end{eqnarray}
where the number without(with) parenthesis of the $\beta_{P^\prime}$ coefficient is the case 
of $\alpha_{P^\prime}=0.500(0.542)$. 
Solving the above equations for $\beta_{P^\prime}$, they are represented by the other 
three parameters as $\beta_{P^\prime}=\beta_{P^\prime}(c_2,c_1,c_0)$, and  
the fitting parameters are $c_2,c_1,c_0,\beta_V,F^{(+)}(0)$ for respective processes.

\begin{figure*}
\resizebox{0.75\textwidth}{!}{%
\includegraphics{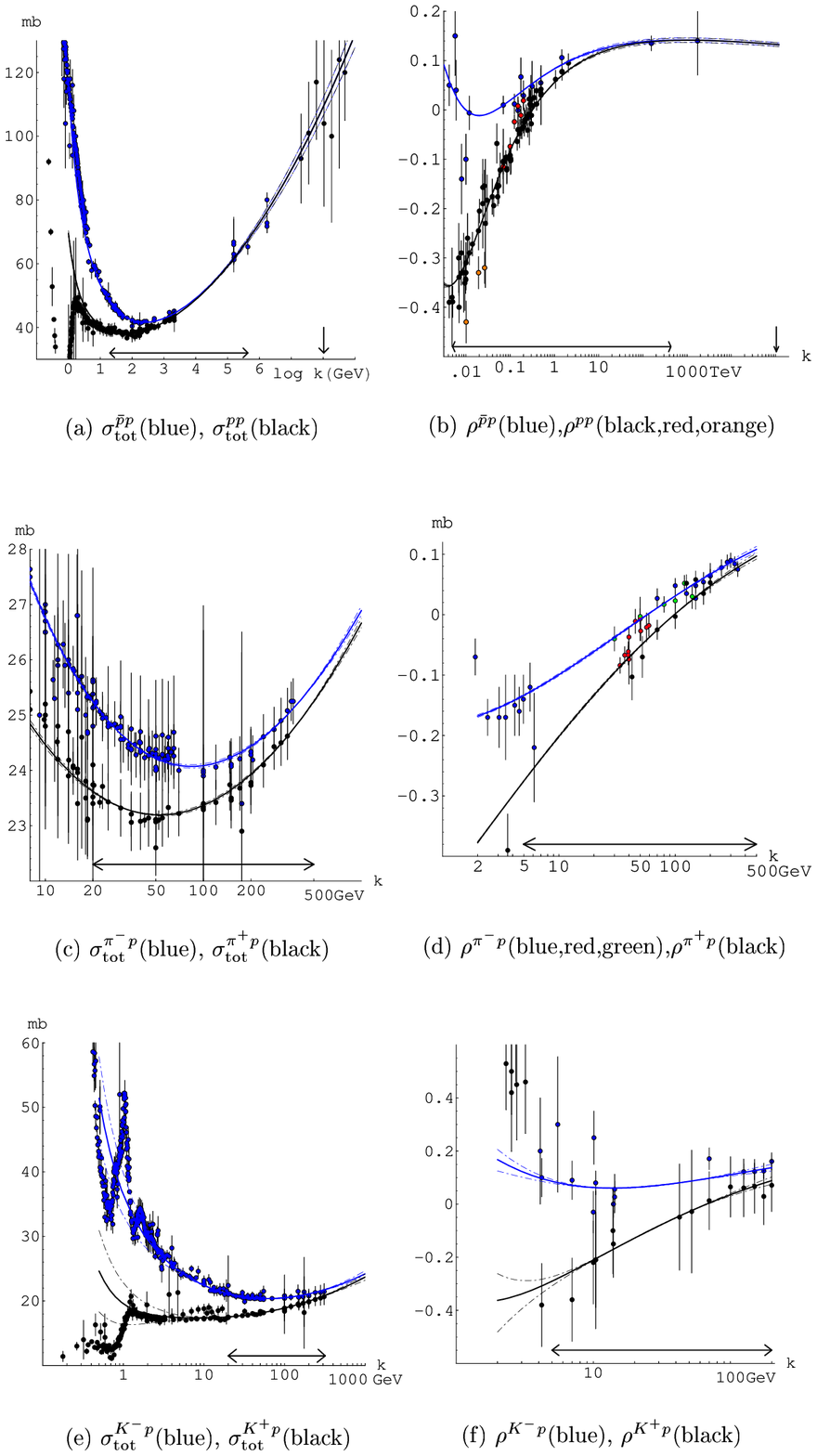}
}% Here is how to import EPS art
\caption{\label{fig1} Results of the fits to (a) $\sigma^{\bar pp,pp}$, (b) $\rho^{\bar pp,pp}$,
(c) $\sigma^{\pi^\mp p}$, (d) $\rho^{\pi^\mp p}$,
(e) $\sigma^{K^\mp p}$, (f) $\rho^{K^\mp p}$. 
%Colors in each caption help to discriminate data points of different groups and processes.
In (b), red(orange) points Fajardo 80\cite{Fajardo80}(Bellettini 65\cite{Bellettini65}) 
of $\rho^{pp}$. 
%Blue points represent $\rho^{\bar pp}$ data. 
In (d), red(green) points are Apokin\cite{Apokin76}(Burq 78\cite{Burq78}) and blue points
are the others in $\rho^{\pi^- p}$ data.
The input-energy regions are shown by horizontal arrows. LHC energy($\sqrt s$=14TeV)
is shown by vertical arrows in (a),(b).
}
\end{figure*}

The result of the fits are depicted in Fig.\ref{fig1} (a)(b) for $\bar p(p)p$ scattering,
(c)(d) for $\pi p$ scattering and (e)(f) for $Kp$ scattering, respectively.  
The values of parameters and $\chi^2$ in the best fits are summarized in Table \ref{tab1}.

There are several comments in the analyses:\\
The fit to the original $\bar pp,pp$ data in Particle Data Group 2006\cite{[9]}
gives the total $\chi^2/(N_D-N_P)=224.8/(240-5)$, where fit to $\rho^{pp}$ data
is unsuccessful, $\chi^{2,\rho}_{pp}/N^\rho_{pp}=126.3/79$, reflecting the situation that
the $\rho^{pp}$ data are mutually inconsistent with different experiments. In Fig.~\ref{fig1}(b)
Fajardo 80\cite{Fajardo80}(red points) and Bellettini 65\cite{Bellettini65}(orange points)
have comparatively small errors, and seem to be inconsistent with the other points by inspection.
We have tried to fit the data set only including Fajardo 80 for $\rho^{pp}$ 
in the relevant energy region, but it is not successful.
We remove these two data from our fit given in Table \ref{tab1}.\\
Similar situation occurs for $\rho^{\pi^- p}$. In Fig.~\ref{fig1}(d)
Apokin 76,75B,78\cite{Apokin76}(red points) in 30GeV$\le k \le 60$GeV, which  
have small errors, are inclined to give smaller values (which are almost in the region of
$\rho^{\pi^+ p}$ data!) than the other data, Burq 78\cite{Burq78}(green points).
We try to fit three types of $\rho^{\pi^- p}$ data:(i) fit to the original data, (ii) fit to 
the data excluding Apokin 76,75B,78 (named Burq fit), and (iii) 
fit to the data excluding Burq 78 (named Apokin fit). 
The fit (i) is almost the same as the Apokin fit( fit (iii)),
however, the $\rho^{\pi^- p}$ data 
around $k=5$GeV are not reproduced well in these fits. 
While they are well described in the Burq fit
(fit (ii)), which is shown by blue line in Fig.~\ref{fig1}(d)). 
Total $\chi^2$ gives respectively 73.82$/$(162-5), 42.25$/$(151-5), 69.8$/$(156-5),
all of which are seemingly successful. However, 
$\chi^{2,\rho}_{\pi^- p}/N_{\pi^- p}^\rho$ are  42.4$/$33, 11.7$/$22, 37.8$/$27, and thus,
only Burq fit is successful. So we adopt the result of Burq fit in Table \ref{tab1}.

\begin{table}
\caption{Values of $B$ parameters in unit of mb. The resluts are given in two cases 
$\alpha_{P^\prime}=0.500,0.542$.}
\begin{center}
\begin{tabular}{lc|c|c}
process  & $B$ & $\alpha_{P^\prime}=0.500$ & $\alpha_{P^\prime}=0.542$\\
\hline
$\bar pp,pp$ & $B_{pp}$ & 0.289$\pm$0.023 & 0.268$\pm$0.024\\
$\pi^\mp p$ & $B_{\pi p}$ & 0.351$\pm$0.036 & 0.333$\pm$0.039\\
$K^\mp p$ & $B_{Kp}$ & 0.37$\pm$0.21 & 0.37$\pm$0.22\\
\hline
\end{tabular}
\end{center}
\label{tab2}
\end{table}

By using Eq.~(\ref{eq10}), we can derive the $B$ parameters from $c_2$ in Table~\ref{tab1}.
The result is given in Table \ref{tab2} in two cases $\alpha_{P^\prime}=0.500,0.542$. 
As seen in Table \ref{tab2}, $B_{pp}$ is somewhat smaller than the $B_{\pi p}$, but 
is consistent within two standard deviation,
although its central value changes slightly depending upon 
the choice of $\alpha_{P^\prime}$. 
Central value of $B_{Kp}$ is consistent with $B_{\pi p}$, 
although its error is very large, due to the present situation of $Kp$ data.
Based on these result, present experimental data are consistent with 
the hypothesis of the universal rise of the total cross section in super-high energies.
On the other hand, $\sigma_{\pi N}^{\rm as}\sim 2/3\ \sigma_{NN}^{\rm as}$\cite{[8]} 
appears
not to be favoured in our analysis.
This is our main result.\\

\noindent \underline{\it Remarks on the analysis of $\pi p$}

In order to obtain the above conclusion,  
it is essential to determine $c_2$ in $\pi p$(or $B_{\pi p}$) with enough accuracy. 
However,
it is very difficult task, since 
the experimental $\sigma^{\pi p}_{\rm tot}$ are reported only in very limited regions 
with momenta $k<400$GeV, in contrast with the  
$\sigma_{\rm tot}^{\bar pp}$ data obtained up to $k = 1.7266\cdot 10^6$GeV.
Actually,
if we fit the same data in the fit of Table \ref{tab1}, using 6 (not 5) parameters  
%($\sigma_{\rm tot}^{\pi^\mp p}$ in $k\ge 20$GeV 
%and $\rho^{\pi^\mp p}\ge 4.95$GeV) 
with no use of the FESR, Eq.~(\ref{eq12}),
we obtain 
\begin{eqnarray}
c_2=(120\pm 46)\times 10^{-5}  &\rightarrow & B_{\pi p}= 0.301 \pm 0.116\ {\rm mb},\ \ \ 
\label{eq13A}
\end{eqnarray}
where $(\alpha_{P^\prime},\alpha_V)=(0.500,0.497)$.
The above value is consistent with the one given in Table~\ref{tab2}, 
$B_{\pi p}=0.351\pm 0.036$mb, within its large error.
However, this error is very large, and the $B_{\pi p}$ in Eq.~(\ref{eq13A})
is consistent with both $B_{pp}$(=0.289mb) and $2/3\ B_{pp}$(=0.193mb).
So by using this value we cannot obtain any definite conclusion.
In other words,
{\it by including the rich informations of the low-energy $\pi p$ scattering data through FESR,
the error of $B_{\pi p}$ is reduced to be less than one third(0.116mb$\rightarrow$0.036mb), 
and as a result,
the universality of $B$ ($B_{pp}=B_{\pi p}$) appears to be preferred.}

In our analysis of Table \ref{tab1},   
$\sigma_{\rm tot}^{\pi^\mp p}$ in $k\ge \overline{N}_2$ 
and $\rho^{\pi^\mp p}$ in $k\ge 4.95$GeV were fitted simultaneously,
using 
FESR(1)(0-$\overline{N}_2$) with $\overline{N}_2=20$GeV.
When we analyze the data by taking $\overline{N}_2=25(30)$GeV,
the $B_{\pi p}$ are determined as
$0.315\pm 0.052(0.303\pm 0.060)$mb.
The results are not so sensitive to the choice of $\overline{N}_2$, 
although their errors become slightly larger.

If we use the FESR(1)(10-20GeV)( not (0-20GeV) ) also for $\pi p$, 
similarly to $\bar pp(pp)$ and $Kp$ and fit the same data,
we obtain $B_{\pi p}=0.314\pm 0.075$mb, which is consistent with our result given in Table \ref{tab2}
but its error becomes about twice the larger of the value in Table \ref{tab2}. 
In order to obtain sufficiently small error of $B_{\pi p}$ 
it appears to be important
to include the informations of low-energy scattering data with 
$0\le k \le 10$GeV through FESR.\\

%In case $\sigma_{\rm tot}^{\pi^\mp p}$ in $k\ge 70$GeV 
%and $\rho^{\pi^\mp p}$ in $k\ge 4.95(70)$GeV are fitted using the same FESR(0-20GeV), 
%the best fitted $B_{\pi p}$ are
%$B_{\pi p}=0.237\pm 0.092(0.206 \pm 0.145)$mb.  Their central values become small,
%although they are consistent with the result in Table \ref{tab2}
%by considering their large errors.\\

%In summary, we can obtain our main result of the universality, because of
%the inclusion of the rich informations of low energy data through FESR.   

Finally we would like to add several remarks:\\
(i) Our $B_{pp}$ , $B_{pp}=0.289\pm 0.023$ mb (in case $\alpha_{P^\prime}=0.500$), 
is consistent with the value of $B$ by COMPETE collab.\cite{[9]}, 
$0.308\pm 0.010$mb, which is obtained 
by assuming the universality of $B$ for various processes.\\
(ii) Our $B_{pp}$ is also consistent with the value by Block and Halzen\cite{[8]}, 
$0.2817\pm 0.0064$mb or $0.2792\pm 0.0059$mb (from the $c_2$ parameter 
in Table III of ref.\cite{[8]}).
A present value of our $B_{pp}$ is located between the above two results.\\
%
%Correspondingly to this situation,
(iii) The universality of $B$ parameter has some theoretical basis from QCD\cite{[15]}.\\
(iv) Our predictions of
$\sigma_{\rm tot}^{pp}$ and $\rho^{pp}$  
at LHC energy($\sqrt s$=14TeV) are
\begin{eqnarray}
\sigma_{\rm tot}^{pp} &=& 109.5 \pm 2.8{\rm mb} \ , \ \   
\rho^{pp}             = 0.133 \pm 0.004\ \ .        
\label{eq14}
\end{eqnarray}
This value is consistent with our previous one, 
$\sigma_{\rm tot}=107.1\pm 2.6$ mb, $\rho=0.127\pm 0.004$\cite{[10]},
which was obtained through the analysis based on only the crossing-even amplitude,
using restricted data sets.
The values of Eq.~(\ref{eq14}) is also located between predictions of the relevant two groups,
$\sigma_{\rm tot}^{pp}=111.5\pm 1.2_{\rm syst}\stackrel{+4.1}{\scriptstyle -2.1}_{\rm stat}$ mb, 
$\rho^{pp}=0.1361\pm 0.0015_{\rm syst}\stackrel{+0.0058}{\scriptstyle -0.0025}_{\rm stat}$\cite{[16]}
and 
$\sigma_{\rm tot}^{pp}=107.3\pm 1.2$ mb, $\rho^{pp}=0.132\pm 0.001$\cite{[8]}.\\
%correspondingly to the situation mentioned above. 
(v) The fit to the $\pi p$ data given in Table \ref{tab1} gives the prediction at $k=610$GeV,
$\sigma_{\rm tot}^{\pi^-p}=25.91\pm 0.03$mb 
(in case of $\alpha_{P^\prime}=0.500$),\footnote{
At this energy, $\sigma_{\rm tot}^{\pi^+p}$ is predicted with $25.62\pm 0.03$mb.
The difference  $\sigma_{\rm tot}^{\pi^-p}-\sigma_{\rm tot}^{\pi^+p}\simeq 0.3$mb. 
} 
which is consistent with the recent observetion
by SELEX collaboration, $\sigma_{\rm tot}^{\pi^- N}=26.6\pm 0.9$mb.\cite{[17]} 
(vi) Finally we would like to emphasize the importance of precise measurements of
$\rho$ ratios in $\bar pp,\ pp,\ \pi^\mp p,\ K^\mp p$ scatterings at intermediate energies 
above $k\ge 5$GeV for further investigations of $B$ parameters.\\

%On the other hand, they give much smaller value of $c_2^{\pi p}$
%than our $B_{\pi p}$. 

\noindent{\it Acknowledgements} This work was (in part) supported by JSPS and French
 Ministry of Foreign Affairs
 under the Japan-France Integrated Action Program (SAKURA).

\end{document}